\documentclass[twocolumn,prl,floatfix,superscriptaddress]{revtex4}
\usepackage[latin9]{inputenc}
\usepackage{textcomp}
\usepackage{mathrsfs}
\usepackage{amsmath}
\usepackage{amssymb}

\makeatletter
\@ifundefined{textcolor}{}
{%
 \definecolor{BLACK}{gray}{0}
 \definecolor{WHITE}{gray}{1}
 \definecolor{RED}{rgb}{1,0,0}
 \definecolor{GREEN}{rgb}{0,1,0}
 \definecolor{BLUE}{rgb}{0,0,1}
 \definecolor{CYAN}{cmyk}{1,0,0,0}
 \definecolor{MAGENTA}{cmyk}{0,1,0,0}
 \definecolor{YELLOW}{cmyk}{0,0,1,0}
}

\usepackage{amsfonts}\usepackage{xspace}
\usepackage{epsfig}
\usepackage{supertabular}
  \usepackage{mathrsfs}

\newcommand{\be}{\begin{equation}}
\newcommand{\ee}{\end{equation}}
\newcommand{\beq}{\begin{eqnarray}}
\newcommand{\eeq}{\end{eqnarray}}
\newcommand{\bea}[2]{\be\label{#2}\begin{array}{#1}}
\newcommand{\eea}{\end{array}\ee}

\def\det{\,{\rm det}\, }

\def\({\left(}
\def\){\right)}

\def\11{1\!\! 1}

\def\varpi{t}

\def\bse{\begin{subequations}}
\def\ese{\end{subequations}}

\def\qli2{{\bf E}}

\makeatother

\begin{document}
\title{
{\Large{}{ Emergent Sasaki-Einstein geometry and AdS/CFT}  }}
\author{Robert J. Berman}
\email{robertb@chalmers.se}

\affiliation{Department of Mathematical Sciences, Chalmers University of Technology,
Gothenburg, Sweden}
\author{Tristan C. Collins}
\email{tristanc@mit.edu}

\affiliation{Department of Mathematics, Massachusetts Institute of Technology, Cambridge, MA 02139, USA}
\author{Daniel Persson}
\email{daniel.persson@chalmers.se}

\affiliation{Department of Mathematical Sciences, Chalmers University of Technology,
Gothenburg, Sweden}
\date{August 27, 2020}
\begin{abstract}
\noindent We consider supergravity in five-dimensional Anti-De Sitter
space $AdS_{5}$ with minimal supersymmetry, encoded by a Sasaki-Einstein metric on a five-dimensional compact manifold $M$. Our main result reveals
how the Sasaki-Einstein metric emerges
from a canonical state in the dual CFT, defined by a superconformal
gauge theory in four dimensional Minkowski space $\mathbb{R}^{3,1}$
in the t'Hooft limit where the rank $N$ tends to infinity. We obtain explicit finite $N-$approximations to the Sasaki-Einstein
metric, expressed in terms of a canonical (i.e. background
free) BPS-state on the gauge theory side. We also provide a string
theory interpretation of the BPS-state in question, which sheds new
light on the previously noted intriguing duality of giant gravitons. 
\end{abstract}
\maketitle

\section{Introduction and summary}

\subsection{Background and motivation}

\noindent A central problem in any quantum theory of gravity is to
explain the \emph{emergence} of the classical spacetime geometry in
some limit of a more fundamental, microscopic description of Nature.
The AdS/CFT-correspondence, introduced in \cite{mal}, provides a
framework in which this problem can, in principle, be addressed. The
AdS/CFT correspondence relates a supergravity theory in the five-dimensional
Anti-deSitter space $AdS_{5}$ to a strongly coupled, rank $N$, superconformal
gauge theory on the 4-dimensional flat boundary $\mathbb{R}^{3,1}$
of $AdS_{5}$. This is a \emph{holographic} correspondence since $\mathbb{R}^{3,1}$
is the (conformal) boundary of $AdS_{5}$; hence it relates a \emph{gravitational
theory} in spacetime, to a \emph{conformal field theory} (without
gravity) on the boundary \cite{ho,su,wi}. In particular, the classical
geometry, i.e. the supergravity vaccuum, should therefore emerge from
a particular quantum state of the dual gauge theory. The main aim
of our work is to make this precise by exhibiting a canonical (i.e.
background free) such quantum state $\Psi_{N}$ and by showing that
the supergravity vacuum in question emerges from the probability amplitude
of $\Psi_{N}$ in the t'Hooft limit where the rank $N$ of the gauge
group tends to infinity. The quantum state $\Psi_{N}$ also sheds
 new light on the intriguing duality of giant gravitons \cite{h-h-i},
which appears in the string theory formulation of the strong form
of the AdS/CFT correspondence. 

In the general
setting of minimal supersymmetry the supergravity vacuum is encoded
by a Sasaki-Einstein metric $g_{M}$ on a five-dimensional compact
manifold $M$ \cite{k-w,m-p,a-f-h-s}. On the gauge theory side the $\mathcal{N}=1$ 
superconformal symmetry is encoded by a \emph{complex cone} $Y$ of
six real dimensions. This means that $Y$ is a complex affine algebraic
variety with a unique singular point $y_{0}$ (the tip of the cone),
endowed with a repulsive holomorphic $\mathbb{R}_{>0}$-action,
representing the conformal dilatation symmetry of the gauge theory.
In the AdS/CFT correspondence the compact manifold $M$ on the supergravity
side arises as the\emph{ base }of the complex cone $Y$ of the gauge
theory: 
\begin{equation}
M:=\left(Y-\{y_{0}\}\right)/\mathbb{R}_{>0}.
\end{equation}
The complex cone $(Y,\mathbb{R}_{>0})$ comes with a canonical
holomorphic three-form $\Omega$ which is homogeneous of degree $3.$
Such a space $Y$ is often called a Calabi-Yau cone in the literature
and is usually endowed with a conical Calabi-Yau metric $g_{Y},$
i.e. a Ricci flat conical K\"ahler metric, as well as a radial coordinate
$r,$ arising as the distance to the vertex point $y_{0}$ with respect to the
Calabi-Yau metric $g_{Y}.$ However, in the background free formalism
that we shall stress, the complex cone $Y$ is not, a priori, endowed
with any metric. This is crucial since our aim is to show how a metric\emph{
emerges }from the metric-independent BPS-sector of the gauge theory. 
Recall that BPS-states can be represented by holomorphic polynomial
functions on the  vacuum moduli space of the rank $N$ gauge
theory, whose mesonic branch is given by the symmetric
product $Y^{N}/S_{N}$, where $S_{N}$ denotes the symmetric group
on $N$ elements.
The vacuum moduli space can thus be described in purely complex algebro-geometric
terms \cite{Beren0,f-h-h-z,m-s-y0}.

\subsection{Summary of results}

\noindent The key point of our proposal is the observation that by imposing
all the manifest symmetries of the rank $N$ gauge theory one naturally
arrives at a canonical quantum BPS-state $\Psi_{N}.$ More precisely,
$\Psi_{N}$ can be realized as a wave function on the  vacuum
moduli space $Y^{N}/S_{N}$ and its
amplitude $|\Psi_{N}|^{2}$ induces a measure on $Y^{N}$
\begin{equation}
|\Psi_{N}(y_{1},...,y_{N})|^{2}\left(\Omega\wedge\bar{\Omega}\right)^{\otimes N}
\end{equation}
 which is $S_{N}-$invariant and $\mathbb{R}_{>0}-$invariant
along each factor of $Y^{N}$ (in this notation $\bar{\Omega}$ denotes
the conjugate of $\Omega$ multiplied by $-i$ so that $\Omega\wedge\bar{\Omega}$
is a volume form on $Y).$ 

Somewhat surprisingly the state $\Psi_{N}$
does not appear to have been considered before in the literature. Its explicit expression
is given in Section \ref{subsec:The-canonical-state}, where the relation
to the BPS-sector of the gauge theory is also explained. 

Now, by quotienting
out the conformal $\mathbb{R}_{>0}-$symmetry we arrive at a measure
on $M^{N}/S_{N},$ which after normalization yields a canonical probability
measure. Equivalently, we obtain  a canonical ensemble
of $N$ ``point-particles'' on $M$.  We show that
a Sasaki-Einstein metric $g_{M}$ on $M$ \emph{emerges} from the canonical
ensemble in the large $N-$limit. In a little more detail, our construction goes as follows. First, the volume
form $dV_{M}$ of $g_{M}$ emerges, after which the metric $g_{M}$ can
be recovered from $dV_{M}$ in a standard manner (by simply differentiating
$dV_{M}$ twice). In fact, in the course of this procedure an $N-$dependent
radial function $r_{N}$ on $Y$ naturally appears in an intermediate
step and it induces a ``quantum correction'' $g_{M}^{(N)}$ to the
Sasaki-Einstein metric $g_{M}$ on $M.$ In this way a limiting radial
coordinate $r$ on $Y$ also naturally emerge from the gauge theory
as $N\rightarrow\infty.$

From the perspective of the AdS/CFT correspondence, the radial
coordinate $r$ on $Y$ corresponds to the radial coordinate on $AdS_{5}$
(as recalled in Section \ref{sec:-giant}) and thus our proposal reveals
how the geometry of $AdS_{5},$ transversal to the conformal boundary
$\mathbb{R}^{3,1},$ naturally emerges from the gauge theory. A ``spin-off
effect'' of this procedure is that it also produces (by a kind of
back-reaction) a conical Calabi-Yau metric on $Y,$ namely the cone
over $g_{M},$ whose distance to the vertex point $y_{0}$ is precisely
$r.$

It should be stressed that there are very few cases known where the Sasaki-Einstein
metric $g_{M}$ on $M$ that can be explicitly computed (but see \cite{g-m-s-w}
for a notable family of exceptions). Thus an important feature of
our construction is that it furnishes canonical approximations $g_{M}^{(N)}$
of Sasaki-Einstein metrics, explicitly expressed in terms of purely
algebro-geometric data.

Finally we give a string-theoretic interpretation of the canonical
quantum state $\Psi_{N}.$ Recall that from the string theory perspective,
the rank $N$ gauge theory arises as the low-energy limit of open
strings attached to a stack of $N$ D3-branes placed at the tip of
the cone $Y$ in the ten-dimensional space-time $\mathbb{R}^{3,1}\times Y.$
On the other hand, when the $N$ D3-branes become massive and charged
they also give rise to a classical extremal black brane, whose near-horizon
limit is the dual ten-dimensional space-time $AdS_{5}\times M;$ the
``horizon'' $M$ thus inherits a Sasaki-Einstein metric $g_{M}.$
We argue that the canonical ensemble 
defined by our quantum state $\Psi_{N}$ represents a bound quantum
state of $N$ \emph{dual giant gravitons }\cite{h-h-i,m-s} (corresponding
to D3-branes wrapping a 3-cycle in $AdS_{5}$ and thus to points on
$M).$ On the other hand, $\Psi_{N}$ can alternatively be viewed
as a coherent state of $N$ \emph{giant gravitons} \cite{m-s-t,bea}
(corresponding to D3-branes wrapping 3-cycles in $M$ and thus to points on $AdS_5$), arising as
the locus where the probability density $|\Psi_{N}|^{2}$ blows-up.
This kind of dual behaviour can be viewed as a realization of a suggestion first put forward
in \cite[Section 5]{h-h-i} in order to explain the intriguing duality
of giant gravitons.

\subsection{Relations to previous results}

\noindent Let us conclude this introduction by briefly mentioning some relations
to previous work. First of all, our work is very much in the spirit
of the program for emergent geometry in AdS/CFT initiated by Berenstein \cite{Berenstein:2004kk,beren}.
The main new feature in our work is the appearence of a negative and fractional
power of the Slater determinant in the definition of the state $\Psi_{N}$
(see equation (\ref{coherentstate})) and its $\beta-$deformation. This is crucial in order to obtain
background independence and to see the emergence of the spacetime metric, as explained in Section~\ref{sec:Emergent}.

As explained in
the end of Section~\ref{subsec:Proof-sketch} our approach builds
on the probabilistic approach to K\"ahler-Einstein metrics on Fano manifolds
introduced in \cite{berm1,berm8,berm8 comma 5}, which, in turn, is
motivated by the Yau-Tian-Donaldson (YTD) conjecture for Fano manifolds.
A different connection between the YTD conjecture and AdS/CFT was
first exhibited in \cite{c-x-y} (compare the discussion in Section
\ref{subsec:Stability}).


\section{Background }

\subsection{\label{subsec:AdS/CFT-and-BPS-states}AdS/CFT and BPS-states }

\noindent The AdS/CFT-correspondence is a conjectural
equivalance between supergravity  in $AdS_{5}$ and a strongly coupled $\mathcal{N}=1$ superconformal gauge theory
on the 4-dimensional flat boundary $\mathbb{R}^{3,1}$ of $AdS_{5}$.
On the gravity side the classical supergravity vacuum is encoded by
a Sasaki-Einstein metric on a compact five-dimensional manifold $M,$
while on the gauge theory side the superconformal symmetry is encoded
by a complex cone $Y$ whose compact base is equal to $M$ (some geometric
background is provided in in Section \ref{subsec:Complex-geometric-setup}). 

The complex cone $Y$ encodes the BPS-sector
of the gauge theory. To see this, first recall that the low-energy dynamics of
a general supersymmetric gauge theory is controlled by the\emph{ moduli
space of classical vacua $\mathcal{M}.$ }The space $\mathcal{M}$
may be defined as the critical points, modulo (complex) gauge equivalence,
of the superpotential $W$ appearing in the (UV-)Lagrangian of the
gauge theory \cite{l-t,a-f-h}. Moreover, the \emph{BPS-operators}
of the gauge theory, i.e. the local operators preserving half of
the supercharges, may be represented by holomorphic (polynomial)
functions on \emph{$\mathcal{M}.$ }They thus form a ring (the coordinate
ring), known as the \emph{chiral} ring of the gauge theory, which
is graded by the  $R$-charge. In the superconformal case
the BPS-operators can be viewed as states (by the usual operator-state
correspondence in a CFT). The BPS-states are chiral primary states
and saturate the BPS--bound, i.e. 
\begin{equation}
\Delta=\frac{3}{2}R.\label{BPScondition}
\end{equation}
 where $\Delta$ denotes the conformal dimension (the mass). As a
consequence, the BPS-sector tends to be robust under non-perturbative
corrections and can thus be used to probe the strong-coupling regime
of the gauge theory. 

In the setting of the AdS/CFT correspondence the mesonic branch of
the moduli space of vacua  of the rank $N$ gauge theory is the symmetric
product \cite{f-h-h-z}
\begin{equation}
\mathcal{M}_N=\text{Sym}^{N}Y=Y^{N}/S_{N}.
\end{equation}
 From the string theory perspective this space parametrizes the transverse
positions of $N$ $D3$-branes inside $Y$. For example, in the maximally supersymmetric $SU(N)-$case
the superpotential $W(Z_{1},Z_{2},Z_{3})$ is defined on 3 complex
$N\times N$ matrices $Z_{i}$ (transforming in the adjoint representation)
and $W=Tr\ensuremath{\left(Z_{1}[Z_{2},Z_{3}]\right).}$ Thus the mesonic
vacuum moduli space $\mathcal{M}_{N}$ may be parametrized by the
set $(\mathbb{C}^{3})^{N}/S_{N}$ of joint eigenvalues of $(Z_{1},Z_{2},Z_{3}),$
showing that, indeed, $Y=\mathbb{C}^{3}$ in this case. Generic single-trace
BPS-operators are given by: 
\begin{equation}
\text{Tr}(Z_{1}^{m_{1}}Z_{2}^{m_{2}}Z_{3}^{m_{3}}),\qquad\Delta=m_{1}+m_{2}+m_{3},
\end{equation}
where $\Delta$ is the conformal dimension. If we also include multitrace
operators one obtains monomials in the eigenvalues of the three operators
$Z_{1},Z_{2},Z_{3}$ which span the mesonic chiral ring of the 
vacuum moduli space. Hence the mesonic (scalar)
BPS-sector in the maximally supersymmetric case is indeed isomorphic
to the ring of holomorphic (polynomial) functions on $(\mathbb{C}^{3})^{N}/S_{N}.$
Gauge theories with minimal supersymmetries may be constructed using
quivers (notably toric ones), encoding the matter content of the Lagrangian,
as well as the superpotential \cite{ken}. 

\subsection{\label{subsec:Complex-geometric-setup}Complex geometric setup }

\noindent In this section we provide complex-geometric background,
emphasizing a background free (i.e. metric independent) perspective
(see \cite{m-s-y} and the monograph \cite{b-g} for the more standard
metric dependent point of view). Let $Y$ be a three-dimensional complex
algebraic affine variety with an isolated singularity $y_{0},$ i.e.
$Y$ may be realized as the zero-locus of a bunch of holomorphic polynomials
on complex space $\mathbb{C}^{M}$ for some positive integer $M.$
Denote by $J$ the induced complex structure on the regular locus
$Y-\{y_{0}\}.$ Assume that $Y$ is endowed with a
\begin{itemize}
\item A repulsive holomorphic $\mathbb{R}_{>0}$-action that fixes $y_{0}$ 
\item a holomorphic top form $\Omega,$ defined on the non-singular locus
$Y-\{y_{0}\},$ which is homogeneuous of degree $3$ with respect to the $\mathbb{R}_{>0}$-action
on $Y.$
\end{itemize}
Such a space $Y$ will here simply be called called a \emph{complex
cone }(but is usually called a Calabi-Yau cone in the physics literature
and an affine Gorenstein cone in the mathenatics litterature). The
form $\Omega$ is uniquely determined up to a multiplicative constant
and can usually be written down explicitely.

\emph{ } For example, in the non-singular
case $Y=\mathbb{C}^{3}$ and $\Omega=dz_{1}\wedge dz_{2}\wedge dz_{3}$
and a large (infinite!) class of singular cases may be obtained by
taking $Y$ to be the hypersurface cut out by a generic quasi-homogenuous
polynomial $F$ in $\mathbb{C}^{4}$ of degree $d.$ defining a log-terminal singularity. Explicitly, that is to say
that there exist positive weights $a_{1},..a_{3}$ such that $F$
is homogeneous of degree $d$ with respect to the corresponding $\mathbb{R}_{>0}$-action:
\begin{eqnarray}
Y&=&\left\{ F=0\right\} \subset\mathbb{C}^{4}
\nonumber \\
F(c^{a_{1}}z_{1},...,c^{a_{4}}z_{4})&=&c^{d}F(z_{1},...,z_{4})\label{eq:hypersurf}
\end{eqnarray}
for any $c\in\mathbb{R}_{>0}.$ In this case $\Omega$ is the
restriction to $Y$ of the three-form $dz_{1}\wedge\cdots\wedge dz_{4}/dF$
and has homogeneuous degree 3 under the assumption that $\sum_{i}a_{i}-d=3,$ which is
the log-terminal condition.

The vector field on $Y$ generating the $\mathbb{R}_{>0}$-action
will be called the \emph{dilatation vector field }and denoted by $\delta.$
Rotating $\delta$ with the complex structure $J$ on $Y$ yields
another vector field that we shall denote by $\xi:$ 
\begin{equation}
\xi=J\delta.\label{eq:xi is J d}
\end{equation}
The space $\mathcal{O}(Y)$ of holomorphic functions on $Y$ decomposes
with respect to the $\mathbb{R}_{>0}$-action: 
\begin{equation}
\mathcal{O}(Y)=\bigoplus_{k=0,1,...}\mathcal{O}_{\lambda_{i}}(Y),\,\,\,\,0=\lambda_{0}\leq\lambda_{1}\leq\cdots,\label{eq:decompositionY}
\end{equation}
where $\mathcal{O}_{\lambda_{k}}(Y)$ is the vector space of holomorphic
(polynomial) functions which are homogeneous of degree $\lambda_{k}$
with respect to $\mathbb{R}_{>0}$. From the perspective of the
underlying superconformal gauge theory the infinitesimal $\mathbb{R}_{>0}$-action $\delta$ represents the conformal dilatation and $\frac{2}{3}\xi$
represents the R-symmetry (the factor $2/3$ ensures that $\Omega$
has the same R-charge, $2,$ as the chiral superspace volume form
$d^{2}\theta,$ where $\theta$ denotes fermionic coordinates of positive
chirality.). Thus equation~(\ref{eq:xi is J d}) is the complex-geometric
realization of the BPS-relation (\ref{BPScondition}) and the summands
$\mathcal{O}_{\lambda_{i}}(Y)$ in the decomposition (\ref{eq:decompositionY})
are thus BPS-states of dimension and $R-$charge equal to $\lambda_{i}$. 

The \emph{base} of a complex cone $Y$ is the compact five-dimensional
manifold $M$ defined by 
\begin{equation}
M:=\left(Y-\{y_{0}\}\right)/\mathbb{R}_{>0}.
\end{equation}
 Thus $M$ is the base of the fibration $(Y-\{y_{0}\})\rightarrow M$
obtained by quotienting out the $\mathbb{R}_{>0}-$action on $Y.$
Since the vector field $\xi$ on $Y$ commutes with the generator
of the $\mathbb{R}_{>0}-$action it induces a vector field on
$M$ that we denote by the same symbol $\xi$ - known as the \emph{Reeb
vector field} on $M$ in mathematics literature. A metric $g_{M}$
on $M$ is said to define a \emph{Sasaki-Einstein metric} on $(M,\xi)$\emph{
}if $g_{M}$ has constant Ricci curvature (normalized so that it
coincides with the Ricci curvature of the standard round metric on the unit-sphere) and is compatible
with the complex structure on $Y.$ 

The compatibility in question can be formulated
in various ways. The crucial point, for our purposes, is that
a Sasaki-Einstein metric $g_{M}$ can be explicitly recovered from
its volume form $dV_{g_{M}}$ as follows. First observe that $dV_{g_{M}}$
induces a\emph{ radial function}, i.e. positive function $r$ on $Y$ which
is one-homogenuous with respect to the $\mathbb{R}_{>0}$-action: 
\begin{equation}
r:=\left(\frac{\iota_{\delta}\Omega\wedge\bar{\Omega}}{dV_{g_{M}}}\right)^{1/6},
\end{equation}
 where we have identified the volume form $dV_{g_{M}}$ on $M$ with
its pull-back to $Y.$ The metric $g_{M}$ on $M$ with volume form
$dV_{g_{M}}$ is a Sasaki-Einstein metric iff the corresponding radial
function $r$ on $Y$ solves the following PDE, after perhaps rescaling
$r,$ 
\begin{equation}
\left(dd^{c}(r^{2})\right)^{3}=\Omega\wedge\overline{\Omega},\,\:d^{c}:=J^{*}d\label{eq:CY eq}
\end{equation}
on the regular locus $Y-\{y_{0}\}$ of $Y$. Here, $d$ denotes the
exterior derivative and $d^{c}$ denotes its ``rotation'' by the
complex structure $J,$ so that $dd^{c}(r^{2})$ defines a two-form
on $Y$ and thus the three-fold exterior product $\left(dd^{c}(r^{2})\right)^{3}$
defines a six-form on $Y.$ 

The PDE (\ref{eq:CY eq}) is the celebrated
\emph{Calabi-Yau equation} on the complex cone $Y;$ it is equivalent
to the condition that the conical K\"ahler metric 
\begin{equation}
g_{Y}:=dd^{c}(r^{2})(\cdot,J\cdot),\label{eq:def of g Y}
\end{equation}
on $Y$ is a\emph{ Calabi-Yau metric} $r^{2},$ i.e. the Ricci curvature
of $g_{Y}$ vanishes. Finally, $g_{M}$ may be explicitely recovered
by identifying $M$ with the level set $\{r=1\}$ in $Y$ and letting
$g_{M}$ be the restrition of $g_{Y}$ to the level set $\{r=1\}.$ 

For example, in the non-singular case when $Y=\mathbb{C}^{3}$ 
the base $M$ is diffeomorphic to the five-sphere $S^5$. The volume form
$dV_{g_{M}}$ of the standard round metric $g_{M}$ on $S^5$
is the unique invariant probability measure. Thus,
by the standard expression for the Euclidean volume form $\Omega\wedge\bar{\Omega}$
in polar coordinates, $r$ is the ordinary radial function on $\mathbb{C}^{3}.$
Accordingly, $g_{Y}$ is the Euclidan metric on $\mathbb{C}^{3}$
and its restriction to the unit-sphere $\{r=1\}$ in $\mathbb{C}^{3}$
indeed recovers the round metric $g_{M}$ on the $S^5$. 

\subsection{\label{subsec:Stability}Stability}

\noindent It is important to emphasize that, in general, the base $(M,\xi)$
of a complex cone $(Y,\mathbb{R}_{>0})$ may\emph{ not} admit
a Sasaki-Einstein metric \cite{m-s-y0,m-s-y,C-S}. Equivalently, this
means that a complex cone may not admit a conical Calabi-Yau metric
and hence no radial solution $r$ to the Calabi-Yau equation (\ref{eq:CY eq}).
Indeed, by \cite{C-S} there exists a Sasaki-Einstein metric on $(M,\xi)$
iff the complex cone $(Y,\mathbb{R}_{>0})$ is\emph{ K-stable.
}This is a purely algebro-geometric condition. For example, if, in
the hypersurface case (\ref{eq:hypersurf}) one takes $Y$ to be given
by
\begin{equation}
\left\{ z_{1}z_{2}+z_{3}^{p}+z_{4}^{q}=0\right\} \subset\mathbb{C}^{4}
\end{equation}
 for given integers $p,q>1,$ then $Y$ admits an $\mathbb{R}_{>0}-$action
with weights $(pq,pq,2q,2p)$ and after rescaling $\mathbb{R}_{>0}$
(to ensure that $\Omega$ has homogeneous degree $3)$ the complex
cone $(Y,\mathbb{R}_{>0})$ is K-stable iff $p/2<q<2p.$ 

As shown
in \cite{c-x-y} the K-stability of $Y$ can be viewed as a generalized
form of the maximization condition for the $a$-central charge of the
SCFT. This means that, in general, there are obstructions to the existence
of a SCFT with a given $\mathbb{R}_{>0}-$graded mesonic chiral
ring $\mathcal{O}(Y^{N}/S_{N}).$ In the present approach a different,
but conjecturally equivalent, stability type condition naturally appears,
which is a variant of the notion of\emph{ Gibbs stability} introduced
in the context of Fano manifolds in \cite{berm8 comma 5}. 

\section{Emergent geometry}
\label{sec:Emergent}

\noindent According to the AdS/CFT-correspondence the classical supergravity
vacuum geometry in $AdS_{5}$ should emerge from some limit of a quantum
state in the dual CFT gauge theory on the boundary. Concretely, in
the present setting the non-trivial part of the supergravity vacuum
in question is encoded by a Sasaki-Einstein metric $g_{M}$ on the
internal compact space $M$, corresponding to the base of the complex
cone $Y$ of the dual gauge theory \cite{m-p,a-f-h-s}. Hence, we
want to show that the Sasaki-Einstein metric $g_{M}$ on $M$ emerges
in a certain ``large $N$-limit'' of a specific (and background
free) quantum BPS-state $\Psi_{N}$ in the dual rank $N$ gauge theory.

\subsection{\label{subsec:The-canonical-state}The canonical state $\Psi_{N}$
and the corresponding probability measure on $M^{N}/S_{N}.$}

\noindent First recall that $Y$ is endowed with a holomorphic top form $\Omega$
and hence one can endow the mesonic classical vacuum moduli space
$Y^{N}/N$ with the volume form $(\Omega\wedge\overline{\Omega})^{\otimes N_{k}}.$
Let $\psi_{1},\dots,\psi_{N_{k}}$ be a maximal number of linearly
independent mesonic BPS-states for the rank $1$ gauge theory with
the same R-charge $\lambda_{k}.$ In other words, $\psi_{1},\dots,\psi_{N_{k}}$
form a basis in the space $\mathcal{O}_{\lambda_{i}}(Y)$ of holomorphic
functions on $Y$ with the same charge $\lambda_{k}.$ Denote by $\Psi_{\text{det}}$
the corresponding Slater determinant, i.e. the totally antisymmetric
holomorphic function on $Y^{N}$ given by 
\begin{equation}
\Psi_{\text{det}}(y_{1},\dots,y_{N_{k}}):=\sum_{\sigma\in S_{N}}(-1)^{|\sigma|}\psi_{\sigma(1)}(y_{1})\cdots\psi_{\sigma(N_{k})}(y_{N_{k}}).
\end{equation}
The function $\Psi_{\text{det}}$ is independent of the choice of
bases in $\mathcal{O}_{\lambda_{i}}(Y)$ up to an overall multiplicative
constant. It thus defines a BPS-state in the rank $N$ gauge theory
(which is baryonic \cite{m-s2,f-h-h-z}). 

We aim to construct a symmetric
measure on the mesonic classical vacuum moduli space $Y^{N}/S_{N}$
which is invariant under the conformal $\mathbb{R}_{>0}-$action
on each factor. One might be tempted to try with the density $|\Psi_{\text{det}}|^{2}$,
but this is unfortunately not $(\mathbb{R}_{>0})^{N_{k}}$-invariant.
The resolution is to simply take a suitable fractional power of the
Slater determinant, namely 
\begin{equation}
\Psi_{N_{k}}:=\Psi_{\text{det}}^{-3/\lambda_{k}}.\label{coherentstate}
\end{equation}
Then, an $(\mathbb{R}_{>0})^{N_{k}}$-invariant measure on $Y^{N}/S_{N}$
is given by the combination 
\begin{equation}
|\Psi_{N_{k}}(y_{1},\cdots,y_{N_{k}})|^{2}(\Omega\wedge\overline{\Omega})^{\otimes N_{k}}.
\end{equation}
This measure is both invariant under the $S_{N}$ permutation symmetry
and the conformal $\mathbb{R}_{>0}$-symmetry, as well as the
R-symmetry. However, since $Y$ is non-compact the integral of this
measure over $Y^{N}/S_{N}$ diverges. To circumvent this problem we
simply quotient by the $(\mathbb{R}_{>0})^{N_{k}}$-action to
get an induced measure on the compact space $M^{N}/S_{N}$. To be
specific, contracting the top-form $\Omega\wedge\overline{\Omega}$
on $Y$ with the dilatation vector field $\delta$ yields a 5-form
$\iota_{\delta}(\Omega\wedge\overline{\Omega})$ on $Y.$ Thus the
$(\mathbb{R}_{>0})^{N_{k}}-$invariant form 
\begin{equation}
|\Psi_{N_{k}}(y_{1},\cdots,y_{N_{k}})|^{2}(\iota_{\delta}\Omega\wedge\overline{\Omega})^{\otimes N_{k}}
\end{equation}
may be identified with a measure on the compact quotient space $M^{N}/S_{N}.$
Finally, the canonical probability measure $d\mathbb{P}_{N_{k}}$
on $M^{N}/S_{N}$ is defined by
\begin{equation}
d\mathbb{P}_{N_{k}}:=\frac{1}{\mathcal{Z}_{N_{k}}}|\Psi_{N_{k}}|^{2}\left(\iota_{\delta}(\Omega\wedge\overline{\Omega})\right){}^{\otimes N_{k}},\label{measureM}
\end{equation}
where 
\begin{equation}
\mathcal{Z}_{N_{k}}:=\int_{M^{N_k}/S_{N_k}}|\Psi_{N_{k}}|^{2}\left(\iota_{\delta}(\Omega\wedge\overline{\Omega})\right){}^{\otimes N_{k}}.\label{partitionfunction}
\end{equation}

Note that in (\ref{measureM}) we have made the implicit assumption
that $\mathcal{Z}_{N_{k}}$ is finite. Since, $|\Psi_{N_{k}}|^{2}$
blows-up along a hypersurface in $Y^{N_k}/S_{N_k}$, namely the zero-locus of
the Slater determinant $\Psi_{\text{det}},$ this is actually a very non-trivial
condition. We intepret it as a consistency condition (which turns
out to be related to the mathematical notions of \emph{K-stability}
and \emph{Gibbs stability}, as elaborated on in \cite{b-c-p}). 

Furthermore,
since $\Psi_{N}$ is holomorphic away from its singularity locus in
$Y^{N}/S_{N}$ (the vanishing locus of $\Psi_{\det})$ the state $\Psi_{N}$
can be viewed as a bound state of BPS-states in the rank $N$ gauge theory (the
role of the singularity locus will be elucidated in Section \ref{sec:-giant}).
As will be elaborated on in \cite{b-c-p} the state $\Psi_{N}$ naturally
lives in a canonical Hilbert space attached to the rank $N$ gauge
theory.

We emphasize that by ``canonical'' we mean that the definition of
$d\mathbb{P}_{N_{k}}$ is \emph{background free}, in the sense that
it does not depend on any underlying metric on $Y$ or $M$. It only
depends on the complex structure $J$ on the classical vaccum moduli space
and the $\mathbb{R}_{>0}$-action and thus only on the superconformal
symmetry of the rank $N-$gauge theory. This is a crucial point for
describing the emergence of the classical Sasaki-Einstein metric $g_{M}$
on $M$, which is our main focus. 

We have also restricted the values of the rank $N$ to be a sequence
of integers 
\begin{equation}
N_{k}:=\dim\mathcal{O}_{\lambda_{k}}(Y),
\end{equation}
i.e.  the multiplicity of the R-charge $\lambda_{k}.$ This
can be seen as a quantization condition. As is well-known, in the
quasi-regular case (discussed in the next section) $N_{k}$ is a polynomial
in $k$ of the form, 
\begin{equation}
N_{k}=\frac{\lambda_{k}^{2}}{2}V+\mathcal{O}(k^{1}),\,\,\lambda_{k}\sim k,\,\,k\rightarrow\infty\label{eq:N in terms of V}
\end{equation}
 where the positive number $V$ is an algebraic invariant of the complex
cone $(Y,\mathbb{R}_{>0}),$ known as its \emph{volume} \cite{m-s-y0,m-s-y,C-S}. 

\subsection{\label{subsec:Main-results}Main results}

\noindent Assume, for simplicity, that the complex cone $(Y,\mathbb{R}_{>0})$
associated to the gauge theory is \emph{quasi-regular}. This is the
generic case; the completely general setting will be treated in the forthcoming work \cite{b-c-p}.

A complex cone $(Y,\mathbb{R}_{>0})$ is \emph{quasi-regular} if (up to a rescaling) the $\mathbb{R}_{>0}-$action
on $Y$ can be complexified to a holomorphic $\mathbb{C}^{\times}-$action.
Denote by $d\mathbb{P}_{N}^{(1)}$ the probability measure on $M$
defined as the {\it $1-$point correlation measure} of the\textit{ }\textit{\emph{canonical
ensemble $(d\mathbb{P}_{N},M^{N}/S_{N})$ introduced in the previous
section. In other words,}} $d\mathbb{P}_{N}^{(1)}$\textit{\emph{
}}is obtained by ``integrating out'' all but one of the factors of $M^{N}:$

\begin{eqnarray}
d\mathbb{P}_{N}^{(1)}(y)&=&\frac{1}{\mathcal{Z}_{N_{k}}}\int_{M^{N-1}/S_{N-1}}|\Psi_{N_{k}}(y,y_{2},...,y_{N})|^{2}
\nonumber \\ 
& &\times  \left(\iota_{\delta}(\Omega\wedge\overline{\Omega})\right){}^{\otimes N_{k}-1}.
\label{eq:def of one pt corre}
\end{eqnarray}
Our main conjecture can now be stated as follows: 

\bigskip{}

\noindent \textbf{Main conjecture:} \textit{Assume that the canonical
ensemble $(d\mathbb{P}_{N},M^{N}/S_{N_{k}})$ is well-defined, i.e.
that $\mathcal{Z}_{N_{k}}<\infty$. Then }
\begin{enumerate}
\item[\textit{(i)}] \textit{The one-point correlation measure $d\mathbb{P}_{N}^{(1)}$
converges, as $N\to\infty$, to the volume form $dV_{M}$ of a Sasaki-Einstein
metric $g_{M}$ on $(M,\xi)$, normalized to have unit-volume, 
\begin{equation}
\lim_{N\to\infty}d\mathbb{P}_{N}^{(1)}=dV_{M};
\end{equation}
}
\item[\textit{(ii)}] \textit{The sequence of radial functions
\begin{equation}
r_{N}:=\left(\frac{d\mathbb{P}_{N}^{(1)}}{\iota_{\delta}(\Omega\wedge\overline{\Omega})}\right)^{-1/6}\label{KahlerpotentialConjecture}
\end{equation}
on the complex cone $(Y, R_{>0})$ converges, as $N\to\infty$, towards
the radial function $r$ of the Calabi-Yau metric $g_{Y}$ on $Y$
corresponding to the Sasaki-Einstein metric $g_{M}$ on $(M,\xi)$. }
\item [\textit{(iii)}]  \textit{Conversely, if there exists a unique Sasaki-Einstein metric
$g_{M}$ on $M,$ then $\mathcal{Z}_{N_{k}}<\infty$. }
\end{enumerate}
We emphasize that for finite $N$, the radial function $r_{N}$ yields
an explicit approximation $g_{M}^{(N)}$ to the Sasaki-Einstein metric
$g_{M}$ on $M$ by identifying $M$ with the level set $\{r_{N}=1\}$
and setting

\begin{equation}
g_{M}^{(N)}=dd^{c}(r_{N}^{2})_{|M}(\cdot,J\cdot)\label{eq:def of g M N}
\end{equation}
(c.f. equation~(\ref{eq:def of g Y})). Hence, part $(ii)$ of the
conjecture is equivalent to 
\begin{equation}
\boxed{\lim_{N\to\infty}g_{M}^{(N)}=g_{M},}\label{convergenceconjecture}
\end{equation}
which is the sought after \emph{emergence} of the Sasaki-Einstein
metric on $M$ in the large $N$ limit.

To be mathematically precise, the convergence statements in the conjecture
are supposed to hold in  the standard weak topologies. In fact,
we make the stronger conjecture that the random measure $N^{-1}\sum_{i=1}^{N}\delta_{x_{i}}$
on the canonical ensemble converges in law towards the deterministic
measure $dV_{M}.$ Below  we will sketch the proof of a \emph{$\beta$-deformed} version of this conjecture. The
complete details of the proof and various elaborations are deferred to the forthcoming work \cite{b-c-p}.

We first introduce a real-analytic family of probability
measures $d\mathbb{P}_{N,\beta}$ on $M^{N}/S_{N}$, defined for a
real parameter $\beta,$ such that $d\mathbb{P}_{N,\beta}$ coincides
with $d\mathbb{P}_{N}$ for $\beta=-1$, if $\mathcal{Z}_{N}<\infty.$
To this end, fix a background radial function $r_{0}$ on $Y.$ We
can then identify the base $M:=\left(Y-\{y_{0}\}\right)/\mathbb{R}_{>0}$
of the cone $Y$ with the level set $\left\{ r_{0}=1\right\} $ in
$Y$ and define $d\mathbb{P}_{N,\beta}$ as follows:
\begin{equation}
d\mathbb{P}_{N,\beta}:=\frac{1}{\mathcal{Z}_{N,\beta}}\left|\Psi_{\det}(y_{1},y_{2},...,y_{N})^{2}\right|^{\frac{3\beta}{\lambda_{k}}}dV_{0}^{\otimes N},\label{eq:def of dP beta}
\end{equation}
 where $dV_{0}$ denotes the volume form on $M$ obtained by restricting
the five-form $\iota_{\delta}\Omega\wedge\bar{\Omega}$ to the level
set $\left\{ r_{0}=1\right\} $ and $\mathcal{Z}_{N,\beta}$ is the
corresponding normalization constant (recall that $N$ is the multiplicity
of the the charge $\lambda_{k})$. The parameter $\beta$ can be viewed
as a regularization parameter, since $\mathcal{Z}_{N,\beta}$ is automatically
finite when $\beta>0$ (or slightly negative). However, it should
be stressed that it is only in the canonical case $\beta=-1$ that
the probability measure (\ref{eq:def of dP beta}) is independent of
the choice of radial function $r_{0}.$ Let $r_{N,\beta}$ be the
radial function on $Y$ defined by
\begin{equation}
r_{N,\beta}:=\left(\frac{d\mathbb{P}_{N,\beta}^{(1)}}{dV_{0}}\right)^{1/6\beta}r_{0}
\end{equation}
(coinciding with $r_{N}$ when $\beta=-1)$ and denote by $g_{M,\beta}^{(N)}$
the corresponding metric on $M,$ obtained by replacing the radial
function $r_{N}$ in equation~(\ref{eq:def of g M N}) with $r_{N,\beta}.$
We then have, in particular:

\bigskip{}

\noindent \textbf{Main Theorem:} \textit{For each fixed $\beta>0$
the following convergence holds
\begin{equation}
\lim_{N\to\infty}r_{N,\beta}=r_{\beta}
\end{equation}
and
\begin{equation}
\lim_{N\to\infty}g_{M,\beta}^{(N)}=g_{M,\beta},\label{convergencetheorem}
\end{equation}
where $r_{\beta}$ is a real-analytic family of radial functions on
$Y$ and $g_{M,\beta}$ is the corresponding real-analytic family
of Sasaki metrics on $M.$ Moreover, if $(M,\xi)$ admits a Sasaki-Einstein
metric, then $r_{\beta}$ and $g_{M,\beta}$ extend real-analytically
to $[-1,\infty[$ and setting $\beta=-1$ yields a Sasaki-Einstein
metric $g_{M}$ on $M.$}

\bigskip{}
In the course of the proof we will show that the square of the limiting radial function
$r_{\beta}$ is the unique conical K\"ahler potential on $Y$ solving
the following PDE on $Y-\{y_{0}\}:$

\begin{equation}
(dd^{c}r_{\beta}^{2})^{3}=\left(\frac{r_{\beta}^{2}}{r_{0}^{2}}\right)^{3(\beta+1)}\Omega\wedge\bar{\Omega}.\label{deformedCYequation}
\end{equation}

\noindent In particular, for $\beta=-1$ this is indeed the Calabi-Yau
equation~(\ref{eq:CY eq}) for the radial function $r$ corresponding
to a Sasaki-Einstein metric $g_{M}$ on $M.$

Loosely speaking, the Main Theorem thus shows that the
Main Conjecture holds after analytic continuation. More precisely,
it shows that the Main Conjecture holds under the assumption that
the order of taking the limits $N_{k}\to\infty$ and $\beta\to-1$
may be interchanged. By a physics level of rigour the Main Conjecture
may thus be considered as established. However, we do expect that
the introduction of the regularization parameter $\beta$ is not needed and, in particular,
that $\mathcal{Z}_{N}<\infty$ if and only if $(M,\xi)$ admits a
unique Sasaki-Einstein metric. As will be shown in \cite{b-c-p},
the ``only if''-direction can be deduced from recent mathematical results
in complex geometry for complex cones $Y$ of any dimension. Proving
the remaining direction appears, however, to be very challenging,
except in the case when $Y$ has complex dimension two, where a direct
proof of the Main Conjecture can be given (as will be shown elsewhere).

The main point of restricting to $\beta>0$ is that the
density of $d\mathbb{P}_{N,\beta}$ is bounded, while for $\beta<0$
it blows up along the hypersurface in $M^{N}$ cut out by the zero
locus $\Psi_{\text{det}}(y_{1},\dots,y_{N_{k}})=0.$ In particular,
for $\beta<0$ the density of $d\mathbb{P}_{N,\beta}$ blows up when
any two points $y_{i}$ and $y_{j}$ on $M$ merge (by the anti-symmetry
of $\Psi_{\text{det}}$). In particular, the canonical ensemble \textit{$(d\mathbb{P}_{N},M^{N}/S_{N})$}
can be viewed as an ensemble of effectively \emph{attractive} points
on $M.$ In Section~\ref{sec:-giant} we will provide a string theory
interpretation of this phenomenon in terms of giant gravitons. But
we first make some comments on the proof of the main theorem (a detailed
mathematical proof will appear in \cite{b-c-p}). 

\subsection{\label{subsec:Proof-sketch}Proof sketch}

\noindent To begin with, we rewrite the probability measure $d\mathbb{P}_{N,\beta}$
as a Boltzmann-Gibbs measure: 
\begin{equation}
\frac{1}{\mathcal{Z}_{N,\beta}}e^{-\beta NE^{(N)}}dV_{0}^{\otimes N},
\label{eq:Gibbs meas}
\end{equation}
where 
\begin{equation}
E^{(N)}(y_{1},...,y_{N}):=-\frac{3}{\lambda_{k}N_{k}}\log|\Psi_{\text{det}}(y_{1},...,y_{N})|^{2}.
\end{equation}
From the perspective of classical equilibrium statistical mechanics
the function $E^{(N)}(y_{1},...,y_{N})$ thus plays the role of the
\emph{energy per particle} and $\beta$ the role of the \emph{inverse
temperature.} Alternatively, $\beta$ can be viewed as a coupling
constant whose sign determines whether the point particles on $M$
are repulsive ($\beta>0)$ or attractive ($\beta<0).$ However, it
should be stressed that this is just a classical statistical mechanical
analogy; the underlying physical system is a quantum one. 

Now, let us make the ``change of variables''
\begin{equation}
\mu(y_1, \cdots, y_N)= N^{-1}\sum_{i=1}^{N}\delta_{y_{i}}
\label{eq:empiric measure}
\end{equation}
 from the space $M^{N}/S_{N}$ to the space $\mathcal{P}(M)$ of probability
measures on $M$. In forthcoming work \cite{b-c-p} we will  show that the following mean field type approximation
holds:
\begin{equation}
E^{(N)}(y_{1},...,y_{N})=E(\mu)+\mathcal{O}(1),\,\,\,N\rightarrow\infty,\label{eq:approx of E N}
\end{equation}
 where $E(\mu)$ is a certain functional on $\mathcal{P}(M)$ with the following property. 
In the case when $\mu$ is $\xi-$invariant, the first variation
of $E$ is given by
\begin{equation}
\delta E(\mu)=-3\varphi_{\mu},\label{eq:first var of E}
\end{equation}
where $\varphi_{\mu}$ is the $\xi-$invariant function on $M$ uniquely
determined (up to an additive constant) as a solution to the following
non-linear PDE on $M$: 
\begin{equation}
\frac{1}{V}(d\eta_{0}+dd^{c}\varphi_{\mu})^{2}\wedge\eta_{0}=\mu.
\end{equation}
 In this equation, $\eta_{0}$ is the (contact) one-form on $M$ obtained by restricting
the one-form $d^{c}\log r_{0}^{2}$ on $Y$ to the level set $\left\{ r_{0}=1\right\} $
and $V$ is the constant defined as the volume of $(Y,\mathbb{R}_{>0})$
(see equation~(\ref{eq:N in terms of V})). For a general $\mu\in\mathcal{P}(M)$
the value of $E(\mu)$ coincides with $E(\bar{\mu})$ where $\bar{\mu}$
denotes the average of $\mu$ over the $\xi-$orbits.

Next, by a general classical argument going back to Boltzmann, the
leading contribution of the volume element $dV_{0}^{\otimes N}$ in
equation~(\ref{eq:Gibbs meas}) is of the form $e^{NS(\mu)},$ where
$S(\mu)$ is the \emph{entropy} of $\mu$ relative to the background
volume form $dV_{0}$ on $M,$
\begin{equation}
S(\mu):=-\int_{M}\log\left(\frac{\mu}{dV_{0}}\right)dV_{0}.
\end{equation}
 Combining the Boltzmann argument with the mean field approximation
(\ref{eq:approx of E N}) we thus arrive at a large$-N$ approximation,
which may be suggestively expressed as
\begin{equation}
d\mathbb{P}_{N,\beta}=e^{-NF_{\beta}(\mu)+o(N)},
\label{eq:free}
\end{equation}
with ``free energy'' given by
\begin{equation}
F_{\beta}(\mu):=\beta E(\mu)-S(\mu)-C_{\beta},
\end{equation}
 where the constant $C_{\beta}$ is the minimum of the functional
$F_{\beta}(\mu)$ on $\mathcal{P}(M).$ 

As will be detailed in \cite{b-c-p},
this approximation can be made mathematically precise using large
deviation theory; the main point is that it implies that the probability
measure $d\mathbb{P}_{N,\beta}$ on $M^{N}/S_{N}$ is exponentially
concentrated on configuration of points $(y_{1},...y_{N})$ such that
the corresponding ``empirical measure'' $\mu$ (equation~(\ref{eq:empiric measure}))
approximates a minimizer $\mu_{\beta}$ of the corresponding free
energy type functional $F_{\beta}$ appearing in equation~(\ref{eq:free}).
Now a direct computation using the first variation (\ref{eq:first var of E})
reveals that a minimizer $\mu_{\beta}$ of $F_{\beta}$ satisfies
the mean field type equation 
\begin{equation}
\frac{1}{V}(d\eta_{0}+dd^{c}\varphi_{\beta})^{2}\wedge\eta_{0}=e^{3\beta\varphi_{\beta}}dV_{0},
\end{equation}
with 
\begin{equation}
\varphi_{\beta}:=\beta^{-1}\log\frac{\mu_{\beta}}{dV_{0}}.
\end{equation}

 Equivalenty, from the perspective of the complex cone $Y,$ this
means that the radial function $r_{\beta}$ on $Y$ defined by 
\begin{equation}
r_{\beta}:=\left(\frac{\mu_{\beta}}{dV_{0}}\right)^{1/6\beta} r_{0},
\end{equation}
satisfies the PDE (\ref{deformedCYequation}), as desired. 

When $\beta>0$
this argument can be made mathematically rigorous and proves the convergence
statements in the main theorem. Finally, if $(M,\xi)$ admits a Sasaki-Einstein
metric, then it follows from essentially well-known results in complex
geometry that there exists a unique family of solutions $r_{\beta}$
to the equation~(\ref{deformedCYequation}), which is real-analytic
in $\beta\in[-1,\infty[$ and converges, as $\beta\rightarrow-1,$
to a solution to the Calabi-Yau equation \eqref{eq:CY eq}. 

Let us also point out that the proof of the mean field approximation
(\ref{eq:approx of E N}), to be detailed in \cite{b-c-p}, exploits
that in the quasi-regular case the compact manifold $M$ is a fibration
\begin{equation}
M\rightarrow X
\end{equation}
of circles over a compact complex space $X,$ which is a Fano orbifold
(the circles are the orbits of the $S^{1}-$action on $M$ induced
from the $\mathbb{C}^{\times}-$action on $Y).$ Moreover, in the
regular case, the canonical probability measure $d\mathbb{P}_{N}$
on $M^{N}/S_N$ may be expressed as the product of the uniform measure
along the $S^{1}-$fibers over $X^{N}$ with the pull-back to $M^{N}/S_N$
of the canonical probability measures on a Fano manifold introduced
in \cite{berm8,berm8 comma 5}. The proof of (\ref{eq:approx of E N})
may then be obtained by generalizing the results on Fano manifolds
in \cite{berm8,berm8 comma 5} to the orbifold setting. 

Interestingly,
from a QFT perspective, the mean field approximation (\ref{eq:approx of E N})
can, using arguments in \cite{berm1}, be viewed as a bosonization
formula with the scalar field $\varphi_{\mu}$ playing the role of
the bosonization of $\psi\bar{\psi}$ when $\psi\in\mathcal{O}_{\lambda_{k}}(Y)$
is viewed as a fermionic operator and $\mu$ is the corresponding
current. 

\section{\label{sec:-giant}$\Psi_{N}$ as a bound state of $N$ (dual) giant
gravitons}

\subsection{BPS brane solutions in AdS/CFT}
\noindent We will now offer a string-theoretic interpretation
of the results established above. First recall that the strong form
of the AdS/CFT correspondence is a conjectural equivalence between
(type IIB) string theory on the ten-dimensional spacetime $AdS_{5}\times M$
and a rank $N$ superconformal gauge theory on the conformal boundary
$\mathbb{R}^{3,1}$ of $AdS_{5}$ \cite{mal,m-p,k-w}. 

The gauge theory
arises as the low-energy limit of open strings in the ten-dimensionsal
space-time $\mathbb{R}^{3,1}\times Y$, whose ends are attached to
a stack of $N$ D3-branes placed at the tip of the cone $Y.$ From
this perspective, the classical vacuum moduli space $Y^{N}/S_{N}$
thus arises as the transverse positions of the $N$ identical D3-branes
in $\mathbb{R}^{3,1}\times Y.$ 

The $N$ coincident $D3$ branes  act as a source of $N$
units of flux of a self-dual five-form $F_{5}$. This induces an extremal
(BPS) charged black brane solution $(g_{l_{N}},F_{5})$ of the supergravity
equations on $\mathbb{R}^{3,1}\times Y,$ with an $N-$dependent large
length scale $l_{N}$ and a horizon at the vertex point $y_{0}$ of
$Y$ (where $l_{N}\sim N^{1/4}l_{P}$ in terms of the Planck scale
$l_{P})$ \cite{h-s}. 

Far a way from $y_{0}$ the internal part of the black
brane-metric $g_{l_{N}}$ is asymptotic to a Calabi-Yau cone metric
$g_{Y}$ on $Y,$ while its near-horizon limit, close to $y_{0},$
is asymptotic to the supergravity metric on $AdS_{5}\times M,$ i.e.
the product of the standard metric $g_{AdS_{5}}$ on $AdS_{5}$ with
a Sasaki-Einstein metric $g_{M}$ on $M.$ In this picture, the radial
coordinate $r$ of the metric 
\begin{equation}
g_{AdS_{5}}=r^{2}g_{\mathbb{R}^{3,1}}+r^{-2}dr\otimes dr,
\end{equation}
in Poincar\'e coordinates, thus corresponds to the distance to the vertex
point $y_{0}$ of $Y$ with respect to the Calabi-Yau metric $g_{Y}$ (in units
where $l_{N}=1$). Moreover, the conformal $\mathbb{R}_{>0}$-symmetry
on $Y$ corresponds to the $\mathbb{R}_{>0}$-isometry on $AdS_{5}.$

\subsection{(Dual) giant gravitons}

\noindent As explained in the previous sections, the canonical wave
function $\Psi_{N_{k}}$ on the classical vacuum moduli space $Y^{N_{k}}/S_{N_{k}}$
represents a BPS-excitation of the vacuum in the gauge theory. Thus,
by the AdS/CFT-correspondence it should be dual to some $N$-particle
state in supergravity on $AdS_{5}\times M$. Which is the relevant
state? We propose that the gravity state corresponding to $\Psi_{N}$
is a \emph{bound state of $N_{k}$ dual giant gravitons}, or, alternatively, a \emph{coherent state of $N$} \emph{giant
gravitons}. Thus the quantum state $\Psi_{N}$ appears to provide a realization of
a suggestion first put forward in \cite[Section 5]{h-h-i}. To explain
this statement we first recall some basic aspects about BPS D3-branes
in this setup.

There are two types of BPS D3-branes on $AdS_{5}\times M.$ On the
one hand we have the \emph{giant gravitons,} which wrap three-dimensional
supersymmetric real submanifolds of $M$ and thus effectively correspond to points on $AdS_5$. These may be realized as
the projection to $M$ of  holomorphic hypersurfaces of $Y$ \cite{bea}. In the present setting we take the holomorphic
hypersurfaces to be of the form $\{\psi=0\}$ for $\psi\in \mathcal{O}_{\lambda_{k}}(Y),$ since we want to describe giant gravitons
with a definite energy. The \textit{dual} \emph{giant gravitons,}
on the other hand, arise from $D3$-branes wrapping an $S^{3}\subset AdS_{5}$
and thus effectively correspond to points on $M$ \cite{m-s}.
It is expected that these two dual descriptions of giant gravitons
should correspond to the same state in the gauge theory \cite{h-h-i}.
Indeed, both the giant gravitons and their duals should be viewed
as excitations of the $N$ background $D3$-branes placed at the tip
of the cone in $\mathbb{R}^{3,1}\times Y.$

We want to make the case that the expected duality of giant gravitons
is naturally incorporated by the canonical quantum state $\Psi_{N}.$
To this end we first recall, following \cite{m-s}, that the quantum
states representing $N$ dual giant gravitons with identical energy
$\lambda_{k}$ can be realized as holomorphic functions on $Y^{N},$
with charge $\lambda_{k}.$ Thus, the probability amplitude $\left|\Psi_{N_{k}}\right|^{2}$
of the state $\Psi_{N_{k}}$ yields a random point process on $M$
with $N$ effectively attractive points, corresponding to a\emph{
statistical ensemble of $N_{k}$ dual giant gravitons on $M$} (with
identical charge $\lambda_{k}).$ On the other hand, $\left|\Psi_{N_{k}}\right|^{2}$
blows up along the locus in $Y^{N}$ where $\Psi_{\text{det}}=0$.
By basic linear algebra, this locus defines a hypersurface in $Y^{N}$
which is made up of all configurations of points $(y_{1},\dots,y_{N_{k}})\in M$
with the property that there exists an element $\psi\in\mathcal{O}_{\lambda_{k}}(Y)$
such that all points $(y_{1},\dots,y_{N_{k}})$ lie on the hypersurface
$\{\psi=0\}$ in $Y$. In other words, 
\begin{equation}
\left|\Psi_{N_{k}}(y_{1},...,y_{N})\right|^{2}=\infty
\end{equation}
precisely when the points $y_{1},...,y_{N}$ are in a very special
position, namely when they are all distributed along the support of
a giant graviton. This means that $\Psi_{N_{k}}$ can alternatively
be interpreted as a \emph{coherent state of $N$ giant gravitons.
} Indeed, in the general formalism of geometric quantization, a coherent
state can be viewed as a quantum state whose probability amplitude
attains its maximum at the corresponding classical configuration in
phase space. 

It should be stressed that the effective attraction between the giant
gravitons is a reflection of the \emph{negative} power appearing in
the definition of $\Psi_{N},$ which in turn, as explained in Section
\ref{subsec:The-canonical-state}, is enforced by conformal invariance.
In fact, by construction, $\Psi_{N}\Omega$ is invariant under the
$\mathbb{R}_{>0}$-action, as well as the R-symmetry. As a consequence,
the state $\Psi_{N}$ has negative R-charge along each factor of $Y^{N},$
which is reminiscent of the appearence of anti-branes and ghost branes
in the matrix models for topological gravity \cite[Section 4.3]{d-w}
and JT-gravity \cite[Section 5.2]{Saad:2019lba}.

Let us conclude by noting that the consistency assumption $\mathcal{Z}_{N}<\infty$
acquires a rather suggestive intepretation in terms of D3 branes/giant
gravitons. Since $\mathcal{Z}_{N}<\infty$ means that the state $\Psi_{N}$
has finite $L^{2}-$norm (once the conformal $\mathbb{R}_{>0}$-symmetry
has been moded out) it translates into the condition that one can
form a \emph{bound} state of $N$ dual giant gravitons on $M^{N}.$
This means that if $\mathcal{Z}_{N}=\infty,$ then $\Psi_{N}$ is
``unstable'' and should decay into a particular configuration of
giant gravitons on $M$ singled out by the property that $|\Psi_{N}|^{2}$
is not integrable in a neighbourhood of the intersection a finite
number $r$ of giant gravitons, $\mathcal{B}_{\psi_{1}}\cap\cdots\cap\mathcal{B}_{\psi_{r}}$
(as will be explained in \cite{b-c-p} these heuristiscs can be made
precise in the mathematical language of log canonical thresholds and
multiplier ideal sheaves). 

In  light of our main theorem this interpretation
is reinforced by the result in \cite{m-t}, which implies that if
$(M,\xi)$ does\emph{ not }admit a Sasaki-Einstein metric, then the
normalized volume forms $dV_{\beta}$ on $M$, appearing in the Main Theorem, converge,
as $\beta\rightarrow\beta_{0}\in [-1,0[$ towards a singular probability
measure on $M$ supported on the intersection $\mathcal{B}_{\psi_{1}}\cap\cdots\cap\mathcal{B}_{\psi_{r}}$
in $M$ for some $\psi_{1},...,\psi_{r}$ in $\mathcal{O}_{\lambda_{k}}(Y).$ 

%


\end{document}